\documentclass[letter,twocolumn,superscriptaddress,showpacs,prb]{revtex4}
\usepackage{graphics,amsmath,graphicx} 
\usepackage{grffile}
\usepackage{color}

\begin{document}

\title{Strong interplay between electron-phonon interaction \\ and disorder in low doped systems}

\author{Domenico Di Sante}
\affiliation{University of L'Aquila, Department of Physical and Chemical Sciences, Via Vetoio, L'Aquila, Italy}
\affiliation{CNR-SPIN, Via Vetoio, L'Aquila, Italy}
\email{domenico.disante@aquila.infn.it}

\author{Sergio Ciuchi}
\affiliation{University of L'Aquila, Department of Physical and Chemical Sciences, Via Vetoio, L'Aquila, Italy}
\affiliation{CNISM Udr L'Aquila}
\affiliation{CNR-ISC, Via dei Taurini, Rome, Italy}
\email{sergio.ciuchi@aquila.infn.it}

\begin{abstract}
The effects of doping on the spectral properties of low doped systems are
investigated by means of Coherent Potential Approximation to describe the
distributed disorder induced by the impurities and Phonon-Phonon Non-Crossing
Approximation to characterize a wide class of electron-phonon interactions which
dominate the low-energy spectral features. When disorder and electron-phonon
interaction work on comparable energy scales, a strong interplay between them
arises, the effect of disorder can no more be described as a mere broadening of
the spectral features and the phonon signatures are still visible despite the
presence of strong disorder.  As a consequence, the disorder-induced
metal-insulator transition, is strongly affected by a weak or moderate
electron-phonon coupling which is found to stabilize the insulating phase.

\end{abstract}

\maketitle

\section{Introduction}

In the last years the developement of more accurate methods of investigations
such as Angular Resolved PhotoEmission Spectroscopy (ARPES), joined with the
fabrication of novel materials as high-Tc superconductors
\cite{Damascelli,Nagaosa}, colossal magnetoresistance manganites \cite{Dagotto},
correlated oxides \cite{Tokura}, topological insulators \cite{Hasan} and
graphene \cite{CastroNeto} in different topological conditions (from bulk, to
surfaces and eterostructures, up to single monolayers), allowed a deep insight
in low-energy electronic and spectral properties. The countinuosly increased
measurements' accuracy in experiments gives the opportunity to detect and study
such low-energy features which in many cases were recognized as the fingerprint
of the electron-phonon interaction.

The possibility to tune the chemical potential by doping offers a great
potentially useful way to modify the materials' electronic structures and
properties. Very recently, in the low doping conditions electron-phonon
signatures were successfully detected in the ARPES spectra of many different
systems, from oxygen vacancies doped $SrTiO_3$ surface
\cite{OxygenVacancesSrTiO3} or lightly bulk doped $SrTiO_3$
\cite{MeevasanaSrTiO3eph,NbDopedSrTiO3}, to monolayer pnictide $FeSe$ growth on 
$SrTiO_3$ \cite{ZXShen}, from tridimensional Anatase \cite{Anatase} to
$Ba_{1-x}K_xBiO_3$ \cite{BaKBiO3} and $Cu_xBi_2Se_3$ \cite{eph-TI}
superconductors, up to $Z_2$ topological non-trivial materials as $Bi_2Se_3$ and
$Bi_2Te_3$ \cite{eph-TI}, as well as on the quasi two-dimensional layered
lightly-doped $Sr_2TiO_4$ \cite{KyleSr2TiO4}. Such a rich variety of different
materials displaying common electron-phonon low-energy features calls for a
deeper understanding of the underlying mechanism at play. However, once all
these systems are taken into account and in particular when dealing with
surfaces, monolayers and low-dimensional systems, the role of disorder cannot be
neglected. In fact the growth processes on substrates and/or the action of
chemical doping imply the presence of disorder, whose impact largely depends on
which energy scale one is focused on. For example, impurity bands can be formed
close to the conduction band of the pristine material as a consequence of  the
presence of the dopant energy levels \cite{KyleSr2TiO4,Si-doped_Ga}, or can have
magnetic origins as in Mn-doped GaAs
\cite{OkabayashiARPESMnDopedSemicondPRB2001,JarrelImpurityBandsMnGaAsPRB2006}. 
On the other hand, oxygen vacancies on the substrate \cite{OxygenVacancesSrTiO3}
may represents centers of scatterings for carriers in the deposited film.
Interestingly in this sense, recent ARPES experiments and ab-initio theoretical
works suggest how charge carriers can be trapped by oxygen vacancies at the
$LaAlO_3$/$SrTiO_3$ (LAO/STO) interface \cite{Claessen} and $SrTiO_3$ surface
\cite{Santander_Syro,Valenti}, naturally introducing the role of disorder in the
understanding of the electronic properties of oxide-oxide heterostructure
interfaces and oxide surfaces, where confined two-dimensional electron gases
(2DEGs) should also undergo superconducting phase transitions \cite{Reyren}.

Usually disorder can be added perturbatively in the theoretical explanation of
ARPES spectra as a weak source of scattering leading to an intrinsic band
linewidth. Within this approach, interactions such as electron-phonon coupling
contribute to the low-energy properties of the spectrum and the disorder simply
provides a further smearing of the electron-phonon features. However this is not the case when
disorder and electron-phonon interaction act on
comparable energy scales. For example let us consider the case of an
intermediate electron-phonon coupling; in the very low doping limit, the system
is prone to polaron formation and the presence of scattering centers may
provide, in a synergic way, the necessary energy to stabilize a small polaron
\cite{Das_Sil,Alexandrov,Mishchenko,Cataudella,Berciu}. Another example is that
of a low (but finite) electron density and weak electron-phonon coupling. In
this case when disorder and electron-phonon interaction are treated
self-consistently impurity and phonon contributions to electron scattering are
not additive when the Fermi energy is of the order of the phonon frequency
\cite{BftPRB2003,KnigavkoCarbottePRB2006}, and  impurity scattering has a
significant nonlinear effect \cite{DoganMarsiglioJSupMagn2007}. In this work we
approach the problem of the interplay between disorder and electron-phonon
interaction starting from a weak electron-phonon coupling, going beyond the
self-consistent Born approximation used in refs.
\cite{BftPRB2003,KnigavkoCarbottePRB2006,DoganMarsiglioJSupMagn2007} by using
the Coherent Potential Approximation (CPA) thus extending our treatement to the
case of strong disorder. Previous studies of models in this peculiar regime
concentrated on the case of classical phonons in binary alloys
\cite{ChenPRB1972} or, in the same context, on the effects
of electron-phonon interaction on transport properties at high
temperature\cite{GirvinPRB1980}. Electron-phonon interaction and strong disorder
have also been studied in the classical phonon case
\cite{LetfulovFreericksPRB2002} within the context of the Falicov-Kimball model of
correlated electrons, for which CPA is the exact solution
\cite{FreericksRMP2003}. Noticeably, the Mott transition in the Falicov-Kimball
model can be described as a disorder-induced metal-insulator transition (MIT) in the alloy
context \cite{FreericksRMP2003}. Here we address the single particle properties,
namely how disorder and electron-phonon interaction modifies ARPES
spectra of lightly-doped materials \cite{note-disorder}. A proper quantum
treatment of the phonon is, in this case, crucial to explain the low energy
features of ARPES spectra. The disorder-induced metal-insulator transition is also
studied as it depends on the strength of the electron-phonon interaction.


The paper is organized as follows: in Section \ref{Hamiltonians} we discuss the
model Hamiltonians and the types of electron-phonon couplings taken into account
in this work.  In Section \ref{Methods} we explain how such models can be solved
in presence of local disorder as introduced by an Anderson type Hamiltonian, and
we discuss the fluctuation of the electron-phonon self-energy due to disorder.
In Section \ref{Results} we present the main results of our work discussing the
interplay between electron-phonon interaction and disorder to explain the
features of the ARPES spectra, we discuss also the electron-phonon dependence of
the disorder-induced metal-insulator transitiion. In Section \ref{Conclusions}
we draw our conclusions and further remarks.

\section{Model Hamiltonians}
\label{Hamiltonians}

We consider in this work an Anderson type Hamiltonian for twodimensional tight-binding
electrons interacting with dispersionless optical phonon modes of the general
form 
\begin{eqnarray}
\label{eq:Hamiltonian1} 
H = H_{el} + H_{ph} + H_{e-ph} + H_{dis} \quad. 
\end{eqnarray}

The electronic nearest-neighbor tight-binding 
part $H_{el} = -t\sum_{<i,j>}(c_i^{\dag}c_j+h.c.)$ gives rise
to a twodimensional energy dispersion $\varepsilon_k=-2t(\cos k_x + \cos k_y)$;
$c_i^{\dag}$ and $c_i$ are the charge carrier creation and annihilation
operators, respectively. The half-bandwidth $D=4t$ will be the energy unit 
throughout the paper and all $k$-vectors are given in units of $\pi/a$
where $a$ is the lattice spacing. We also choose the zero energy level 
$\omega=0$ to the position of the chemical potential.

The disorder part is assumed to be of the Anderson type 
\begin{eqnarray}
\label{eq:AndersonHamiltonian} 
H_{dis} = \sum_{i} \xi_{i} c_i^{\dag}c_i \quad ,
\end{eqnarray} 
where $\xi_i$ are disorder independent random energies taken
according to the following disorder distributions: \\ i) the bimodal $P_i(\xi)=
x\delta(\xi-E_b)+(1-x)\delta(\xi)$ characterizing a concentration of $x$ impurities in the host material, 
\\ ii)  the gaussian $P_g(\xi)=
(1/\sqrt{2\sigma^2})\exp(-\xi^2/2\sigma^2)$ where $\sigma^2$ is the
disorder variance to mimic a conformational disorder, \\ iii) or as the sum of two independent variables, one of
which distributed according to $P_i$, and the other one distributed according
to $P_g$.

For the free phonon part, we assume a simple undispersed Einsteins' phonon
Hamiltonian $H_{ph}=\omega_0\sum_i a^\dagger_i a_i$ with a characteristic phonon
frequency $\omega_0$. We fix the value of the phonon frequency in the adiabatic regime
$\omega_0/D=0.05$.

For the electron-phonon interaction part $H_{e-ph}$ 
we consider three different kinds of models. The first two can be 
obtained from the following
density-displacement Hamiltonian


\begin{equation}
\label{eq:localnonlocaleph}
H_{e-ph}=-\sum_{i,j}g_{i,j} c^\dagger_i c_i (a_j+a^\dagger_j) \quad .
\end{equation}
The Holstein local (LOC) model is obtained when $g_{i,j}=g\delta_{i,j}$,
whereas a general, even long-range Fr\"ohlich type interaction (NLOC), can be
considered in more general cases. In the spirit of our work, here we focus our
attention on the two-dimensional screened Fr\"ohlich type interaction. Let us
consider the long wavelength limit of the Fourier transform of the longitudinal
optic (LO) polar coupling $[g^2]_{i,j}=\sum_k g_{i,k}g_{k,j}$ 
\begin{equation}
g^2({\bf k})=\frac{1}{N}\sum_{R}e^{-i{\bf k}{\bf R}} [g^2]_{i,i+R} \quad . 
\end{equation} 
If $g^2({\bf k})$ is of the Fr\"ohlich type, i.e. $g^2({\bf k})\propto 1/k^2$, after summing
over all possible value of $k_z$, we get an effective coupling which at small $k$ behaves as
$g^2({\bf k})\propto 1/k$ depending only on the two-dimensional wave-vector ${\bf k}$ \,
\cite{DasSarma}. Since in our model electrons are free to have planar motions,
we next consider the action of the two-dimensional screening of the in-plane
carriers. This screening is independent on the carrier density, and the
effective coupling is thus replaced by $g^2({\bf k})\rightarrow
g^2({\bf k})/\epsilon({\bf k},\omega=0)$, where $\epsilon({\bf k},\omega=0)=1+\kappa/k$ and
$\kappa=2m^*e^2/\hbar^2\epsilon_r$ is the two-dimensional screening wave-vector.
The large $k$ behaviour of $g^2({\bf k})$ is obtained restoring the lattice symmetries by
replicating the small $k$ form 
\begin{equation}
\label{eq:Gvect}
g^2({\bf k})= \frac{C}{N_{{\bf G}}} \sum_{\bf G} \frac{|{\bf k}+{\bf G}|}{|{\bf k}+{\bf G}|+\kappa}
\end{equation}
where ${\bf G}$ is a reciprocal lattice vector and $N_{{\bf G}}$ is the number of summed terms in eq. (\ref{eq:Gvect}). 
For our aims, we find that  a summation over the nearest-neighbor reciprocal vectors is sufficient. 
The normalization constant $C$ is chosen by fixing the value of a coupling constant $g$
\begin{equation}
\label{eq:g}
g^2=\frac{1}{N}\sum_{\bf k}g^2({\bf k}) \quad.
\end{equation}
In both LOC and NLOC models the dimensionless electron-phonon coupling constant is defined in terms of $g$ as
\begin{equation}
\label{eq:lambda}
\lambda=2g^2/\omega_0 D\quad.
\end{equation}


Another model which we consider in this work is the so-called interaction with a
phonon mode such as that occurring with apical oxygens in layered
perovskites \cite{BMcuprates}, to which hereafter we refer  as Apical Oxygens Hamiltonian (AO)
\cite{SawatzkyBM}. The form of the Hamiltonian is the same as in
eq. (\ref{eq:Hamiltonian1}), but now we consider several two-dimensional planes
where electron carriers are free to move (index $\alpha$) unconnected by
out-of-plane hopping processes. The interaction between different planes is
introduced through the following AO electron-phonon coupling  
\begin{eqnarray}
\label{eq:BMe-ph}
H^{BM}_{e-ph}=-\frac{g}{\sqrt{2}}\sum_{i.\alpha}c_{i,\alpha}^{\dag}c_{i,\alpha}(x_{i,\alpha+1/2}-x_{i,\alpha-1/2}) \quad,
\end{eqnarray}
where $x_{i,\alpha+1/2}$ is the (dimensionless) displacement
$x_{i,\alpha+1/2}=(a^\dagger_{i,\alpha+1/2}+a_{i,\alpha+1/2})$ of the interplane
apical atom in the $i$-th site of the $\alpha$-th plane. Within this AO model,
disorder variables are chosen uncorrelated as before, and the Anderson term now
reads as $H^{BM}_{dis} = \sum_{i,\alpha} \xi_{i,\alpha}
c_{i,\alpha}^{\dag}c_{i,\alpha}$.
In AO model the dimensionless electron-phonon coupling constant is defined 
as in the LOC and NLOC models through Eq. (\ref{eq:lambda}).

\section{Methods of solution for local and non-local electron-phonon Hamiltonians} 
\label{Methods}

\subsection{CPA and Phonon-Phonon Non-Crossing Approximation in the Holstein model}

Here we introduce our approximations in the case of purely local electron-phonon
interaction (LOC). We use the CPA to treat
the local disorder. The CPA can be thought as an exact theory on an infinite
coordination lattice\cite{Siggia}; for this reason it is therefore much similar
to the single-site Dynamical Mean Field Theory (DMFT) \cite{Vollhardt,DMFT}. As
in DMFT, for solving the LOC model we consider a single site embedded into 
a self-consistent medium \cite{DMFT}. 
The single-site propagator $\mathcal{G}$ can be expressed in terms
of a  local propagator which embodies the {\it average} action of the
environment ($G_{0}(\omega)$) and a self-energy $\Sigma(\omega)$
\cite{DMFT}:
\begin{equation}
\label{eq:Gimp0}
\mathcal{G}(\omega)=\frac{1}{G_{0}^{-1}(\omega)-\Sigma (\omega)}\quad .
\end{equation}

The site propagator $\mathcal{G}$ can be expressed as an average over disorder variable
(hereafter a generic quantity $A$ which depends on disorder realizations is
denoted by $\hat{A}$ while its average is $A=[\hat{A}]_{\xi}$)
\begin{eqnarray}
\label{eq:Gimp}
\mathcal{G}(\omega)=
\left [
\frac{1}{G_{0}^{-1}(\omega)-\xi-\hat{\Sigma}_{eph}(\omega)} \right ]_\xi \quad,
\end{eqnarray}
where $\xi$ is the local disorder variable and 
$\hat{\Sigma}^{eph}(\omega)$ is the electron-phonon self-energy which 
depends on the local disorder variables.

\begin{figure}[htp]
\begin{center}
\includegraphics[width=0.45\textwidth,angle=0,clip=true]{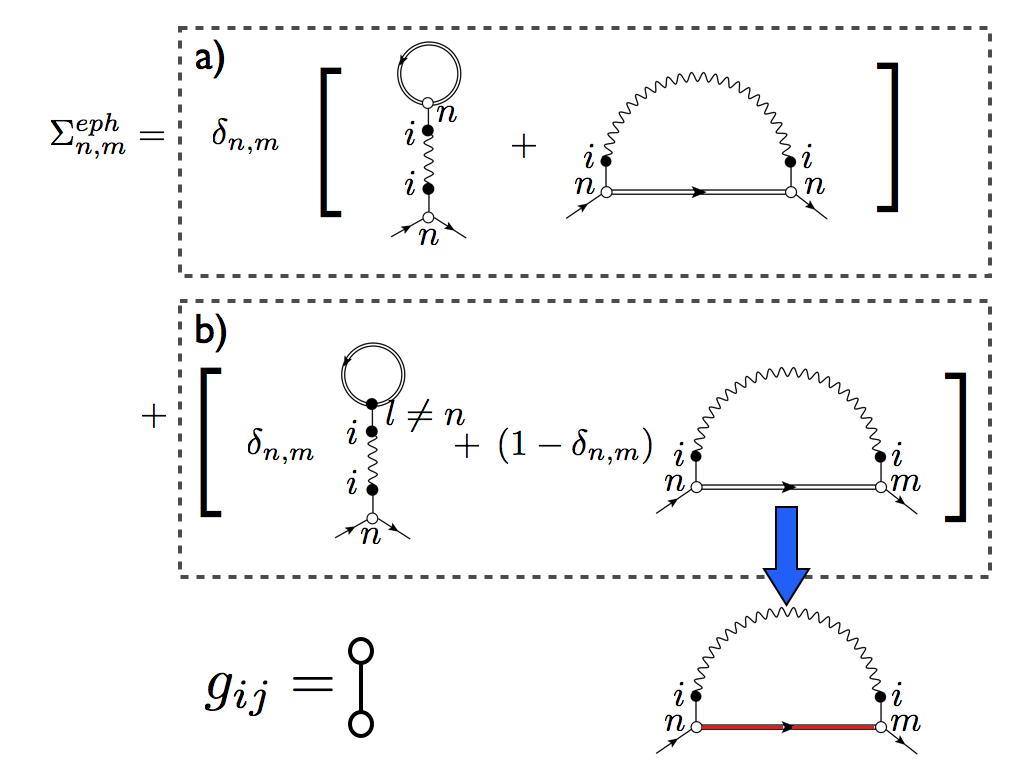}
\caption{Electron-phonon interaction diagrams. Open straight line is the
non-averaged electron propagator, filled straight line is the disorder-averaged
electron propagator, wavy line is the phonon porpagator.}
\label{fig:DiagramsEPH}
\end{center}
\end{figure}

Electron-phonon interaction in the LOC model can be self-consistently taken into
account within a CPA --- or equivalently DMFT --- scheme at zero electron density \cite{Bronold}.
At finite electron density we choose a self-consistent Phonon-Phonon Non-Crossing
Approximation (PPNCA) for the electron-phonon self-energy  
\cite{note-NCA} (see diagrams of type a) in Fig. \ref{fig:DiagramsEPH}):
\begin{eqnarray}
\hat{\Sigma}_{eph}(\omega)&=&-\frac{g^2}{\beta}\sum_{m}D^0(\omega-\imath\omega_m)\hat{\mathcal{G}}(\imath\omega_m) \nonumber \\
                                &+& \hat{\Sigma}_H \quad ,
\end{eqnarray}
where $D^0(\omega)$ is the free-phonon Green's function while the  frequency
independent Hartree term of the electron-phonon self-energy
\begin{equation}
\label{eq:Hartree}
\hat{\Sigma}_H=-\frac{2 g^2}{\omega_0}\hat{n} 
\end{equation}
is expressed in term of the
local density $\hat{n}$ which is given by $\hat{n}=-\frac{1}{\beta}\sum_n
\hat{\mathcal{G}}(\imath\omega_n)e^{i\omega_n 0^+}$. 
In this approximation the phonon propagator is not renormalized by the electron
density fluctuations; we therefore associate the phonon frequency to that
obtained by experiment or assume that the phonon frequency renormalization is
negligible at low electron density. 

After Matsubara's frequency summation the PPNCA self-energy is written as
\begin{eqnarray}
\label{eq:NCA}
\hat{\Sigma}_{eph}(\omega)=g^2\int\,d\epsilon \hat{A}(\epsilon)
\left[\frac{b(\omega_0)+f(\epsilon)}{\omega+\omega_0-\epsilon+\imath\delta}
+\right.\nonumber \\
\left. \frac{b(\omega_0)+1-f(\epsilon)}{\omega-\omega_0-\epsilon+\imath\delta}\right] 
+ \hat{\Sigma}_H \quad ,
\end{eqnarray}
with $b(\omega_0)$ and $f(\epsilon)$ referring to the Bose-Einstein and
Fermi-Dirac distributions respectively, and $\hat{A}(\epsilon)=(-1/\pi)\Im
\hat{\mathcal{G}}(\epsilon)$ being the spectral function.

The averaged propagator is translationally invariant. It can be expressed in
terms of the local self-energy as: 
$G(k,\omega)=1/[\omega-\epsilon_k-\Sigma(\omega)]$. The  averaged local 
propagator is thus:
\begin{eqnarray}
\label{eq:Gloc}
G_{loc}(\omega)=\int\,d\epsilon N(\epsilon)\frac{1}{\omega-\epsilon-\Sigma(\omega)} \quad ,
\end{eqnarray}
where $N(\epsilon)=\sum_k\delta(\epsilon-\epsilon_k)$ is the non-interacting
density of states. The self-consistency condition requires the single-site Green's
function (\ref{eq:Gimp}) to coincide with the local lattice Green's function
(\ref{eq:Gloc})
\begin{equation}
\label{eq:selfcons}
G_{loc}(\omega)=\mathcal{G}(\omega)\quad .
\end{equation}

In this way equations
(\ref{eq:Gimp0},\ref{eq:Gimp},\ref{eq:NCA},\ref{eq:Gloc},\ref{eq:selfcons})
define a self-consistency loop to be iterated to get the self-consistent local
self-energy which takes into account disorder at the CPA level as well
electron-phonon interaction coming only from diagrams of type a) in Fig. \ref{fig:DiagramsEPH}. We call this
scheme PPNCACPA. From the operative point of view, starting with an educate ansatz
for $G_0$, we use Eqs. (\ref{eq:Gimp},\ref{eq:NCA}) to determine $\mathcal{G}$, Eq.
(\ref{eq:selfcons}) to obtain $\Sigma$ and  Eq. (\ref{eq:Gimp0}) to obtain a new
$G_0$ for iterating the procedure.
This iteration scheme differs from DMFT due to the approximate treatment of the 
electron-phonon interaction trough PPNCA.

\subsection{PPNCACPA in the AO model}

\begin{figure}[t]
\begin{center}
\includegraphics[width=0.5\textwidth,angle=0,clip=true]{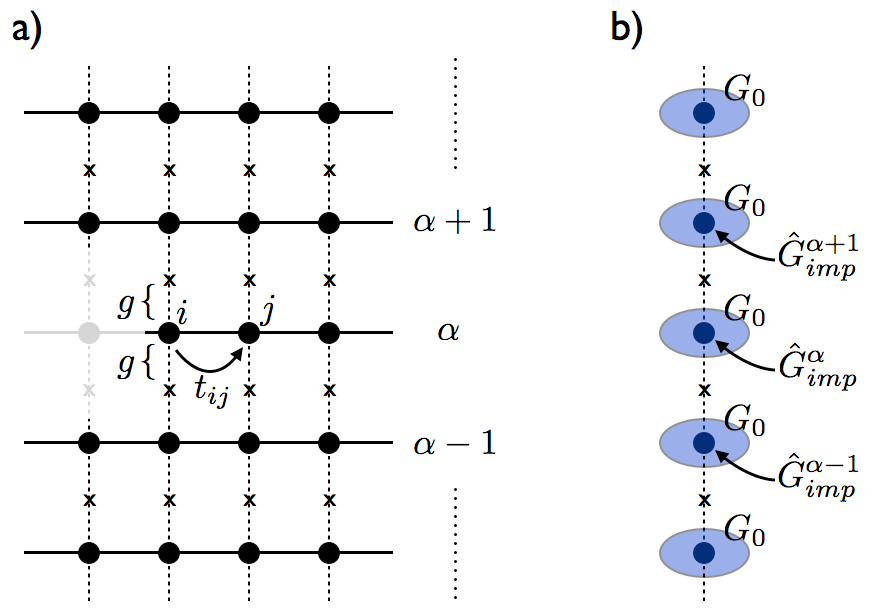}
\caption{ DMFT mapping of the AO model. a) Lattice model in which electrons move on the planes and interact with the AO phonon. 
b) Mapping of the lattice problem into a single chain single-site model.}
\label{fig:DMFTBM}
\end{center}
\end{figure}

To generalize PPNCACPA to the AO model we have to introduce the planar structure
into our single-site model. We have a chain of single-site models as depicted in Fig.
\ref{fig:DMFTBM}. The interaction between neighboring planes occurs through the
electron-phonon interaction (see Eq. (\ref{eq:BMe-ph})). In the AO model we
neglect the interplane hopping and therefore the self-consistent $G_0$ is
plane-diagonal. Eq. (\ref{eq:Gimp}) can be generalized as
\begin{eqnarray}
\label{eq:GimpBM}
\mathcal{G}(\omega)=\left [\frac{1}{G_{0}^{-1}(\omega)-\xi_\alpha-\hat{\Sigma}^{\alpha}_{eph}(\omega)}\right ]_\xi \quad,
\end{eqnarray}
where $\alpha$ is the plane index.
Notice that after averaging $\mathcal{G}$ does not depend on the plane indexes.

Now we have to generalize Eq. (\ref{eq:NCA}) to the BM model.
Defining the upper and lower  local phonon propagators as
\begin{equation}
D^{(\pm)}(t) = -i \langle T x_{i,\alpha\pm 1/2}(t) x_{i,\alpha\pm 1/2}(0) \rangle \quad ,
\end{equation}
the Fock and Hartree terms of the electron-phonon self-energy take the form 
\begin{eqnarray}
\label{eq:NCABMMatsu}
\hat{\Sigma}^\alpha_{F}(\omega)=-\frac{g^2}{2\beta}\sum_{m}D^{+}(\omega-\imath\omega_m)\hat{\mathcal{G}}^{\alpha}(\imath\omega_m) -\nonumber\\
-\frac{g^2}{2\beta}\sum_{m}D^{-}(\omega-\imath\omega_m)\hat{\mathcal{G}}^{\alpha}(\imath\omega_m)\quad ,\\
\hat{\Sigma}^\alpha_{H}=\frac{g^2}{2} \left (D^{+}(0)\hat{n}^{\alpha} - D^-(0)\hat{n}^{\alpha+1} \right ) + \nonumber \\
\frac{g^2}{2}  \left (D^{+}(0)\hat{n}^{\alpha} - D^-(0)\hat{n}^{\alpha-1} \right )\quad ,
\end{eqnarray}
where $D^{(\pm)}(i\omega_n)$ are the local phonon propagators in the Matsubara
frequencies and  $\hat{n}^{\alpha}=-\frac{1}{\beta}\sum_n
\hat{\mathcal{G}}^{\alpha}(\imath\omega_n)e^{i\omega_n 0^+} $ is the local
density on a generic site of the plane $\alpha$. Notice that $\hat{n}^\alpha$ still
depend on the disorder realization. Notice also that interplane coupling occurs
due to the Hartree term in the self-energy Eq. (\ref{eq:NCABMMatsu}). After
Matsubara's frequency summation the Fock contribution to the self-energy is
written as
\begin{eqnarray}
\label{eq:NCABM}
\hat{\Sigma}^\alpha_{F}(\omega)=g^2\int\,d\epsilon \hat{A}^\alpha(\epsilon)\left[\frac{b(\omega_0)+f(\epsilon)}{\omega+\omega_0-\epsilon+\imath\delta}+\right.\nonumber\\
					+\left.\frac{b(\omega_0)+1-f(\epsilon)}{\omega-\omega_0-\epsilon+\imath\delta}\right] \quad ,
\end{eqnarray}
with $\hat{A}^\alpha(\epsilon)=(-1/\pi)\Im \hat{\mathcal{G}}^{\alpha}(\epsilon)$
being the $\alpha$-th plane spectral function. The scheme of iteration is
basically the same as for the Holstein (LOC) model with an important difference:
we have to iterate the self-consistency condition for an array of planes.
Adopting periodic boundary conditions, we need 64 planes to achieve convergence for
the sets of parameters used throughout the paper.

\subsection{Generalization to non-local models of electron-phonon interaction}

Now let us consider a general non-local electron-phonon interaction as that of
the model NLOC Eq. (\ref{eq:localnonlocaleph}). The perturbation theory in terms
of the electron-phonon coupling constant $g_{i,j}$ can be written in the lattice
space. This is shown diagramatically in Fig. \ref{fig:DiagramsEPH}. The diagrams
sets are divided into two groups: a) refers to local type diagrams in which only the
$[g^2]_{n,n}$ appears (see discussion about the LOC model) while b) contains extra terms which include $[g^2]_{n,m}$ for $\;\;m\ne n$.
We divide our calculation into two steps.

In a first step we implement the PPNCACPA previously described for the Holstein
(LOC) model taking into account the a) diagrams for the electron-phonon
interaction. We use in this stage a coupling constant $g^2=[g^2]_{i,i}$. Within
such a treatement, we are taking into account disorder and electron-phonon
interaction at the local level. Now we include the non-local part of
electron-phonon interaction including diagrams of type b) {\it at the average
level}, i.e. we consider the internal propagator averaged over disorder. Average
restores translational invariance and the Hartree term (tadpole diagram in Fig.
\ref{fig:DiagramsEPH}b)), which is independent on frequency, can be reabsorbed
in the definition of the chemical potential. The only relevant term is the Fock
one averaged over disorder, as depicted in Fig. \ref{fig:DiagramsEPH}b) and 
highlighted by the blue arrow. The
self-energy thus takes into account both disorder and electron-phonon
interaction, while disorder and local part of electron-phonon interactions
(diagrams a)) are evaluated self-consistently; the non-local part is taken into
account non-selfconsistently on a final stage. Therefore this approach should
not be extended to the polaronic type of couplings. However, due to the
relevance of disorder in our calculations, we have checked that the results are
quite insensible to the actual value of the screening wavevector provided that $\kappa > 0.001$, 
and thus on the specific form of the non-local e-ph coupling.

\subsection{Alternative CPA schemes}
  
In order to investigate the correlations in the one particle spectra between
disorder and electron-phonon interaction, in the local PPNCACPA loop 
we can compare two CPA
schemes, the one we are actually using in which the electron-phonon self-energy
do depend on local random potentials (CPA2) and a more simpler scheme in which
we average the e-ph self-energy diagrams of type a) on disorder (CPA1). In the
case of NLOC models, to take into account the non-locality of the electron-phonon
interaction, we finally implement the second stage of our approximation having
the local self-energy from CPA2 or CPA1 formulations. Notice that CPA1 scheme in
absence of electron-phonon interaction is usually referred as {\it virtual-crystal}
approximation\cite{virtual}. The comparison between the two schemes sill give us an 
idea of the relevance of the electron-phonon self-energy fluctuations due to disorder at different 
energy scales.

\begin{figure}[htp]
\begin{center}
\includegraphics[width=0.45\textwidth,angle=0,clip=true]{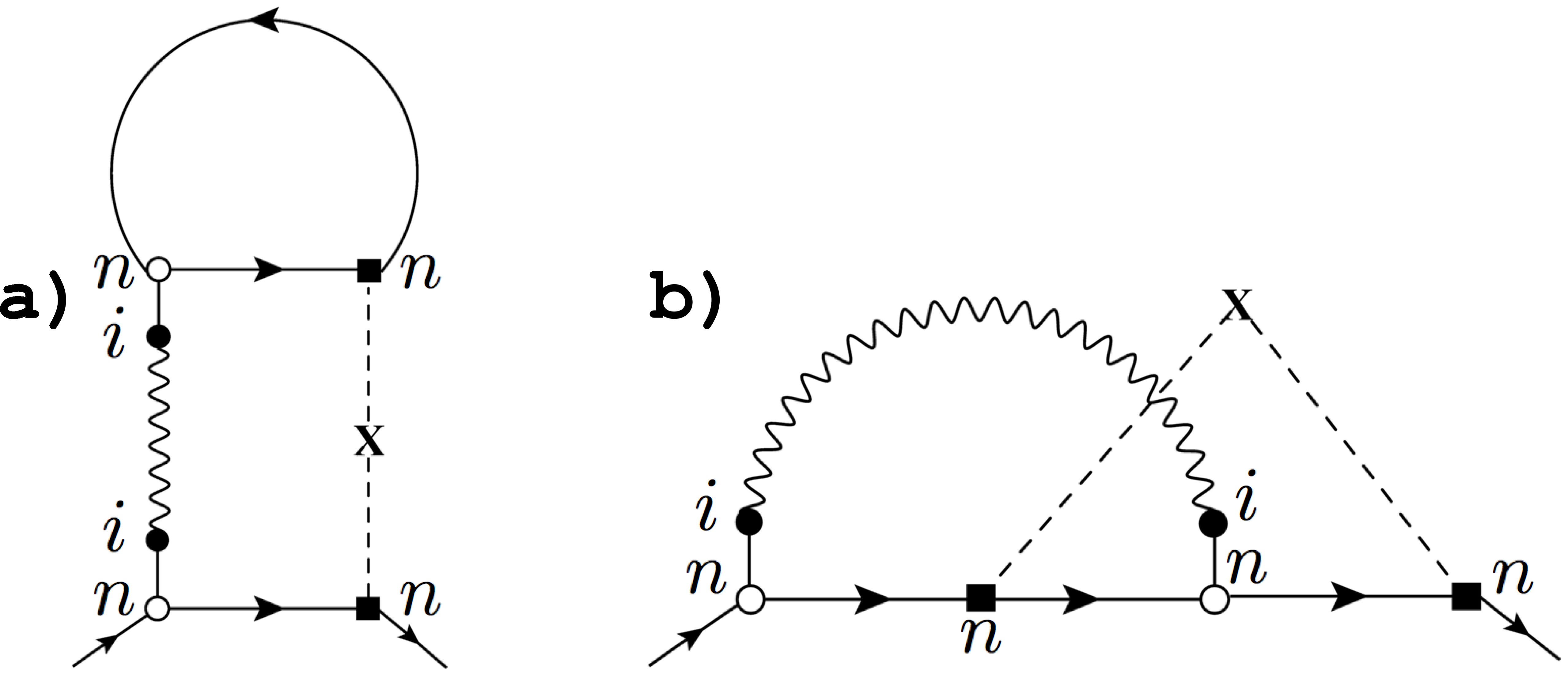}
\caption{Examples of diagrams neglected in CPA1 scheme for gaussian distributed disorder. 
a) A correction which takes into account disorder 
 correlations in the Hartree part of the self-energy entering in Eq. (\ref{eq:Gimp}) within CPA2 
 scheme but neglected in the same expansion within CPA1 scheme. Solid line
 represents the self-consistent propagator $G_0$, wavy line the phonon
 propagator, dashed line disorder insertion. 
b) A disorder-induced vertex correction appearing in
 the expansion of the Fock part of the local electron-phonon self-energy.}
\label{fig:CPA2vsCPA1diagrams} 

\end{center}
\end{figure}

We notice that, averaging the internal propagators appearing in diagrams of type
a) shown in Fig. \ref{fig:DiagramsEPH}, means substituting the internal electron
propagators with their averages. The Hartree contribution (tadpole diagram in
fig. \ref{fig:DiagramsEPH}a)) averages to a frequency and $k$-independent value
thus reducing to a mere shift of the chemical potential. The remaining
contribution is the Fock term in which the internal propagator has been averaged
over disorder. This average procedure neglects i) correlations between the
density and the disorder variable at a given site and ii) disorder and
electron-phonon correlated scatterings. From a perturbative point of view the
diagrams which contribute to these two mechanisms are depicted in Fig.
\ref{fig:CPA2vsCPA1diagrams}.
We notice that, due to our strong-disorder approach, these contributions are not included 
in the self-consistent Born approximation approach of ref. \cite{BftPRB2003}.

\section{Results}
\label{Results}

\begin{figure}[htp]
\begin{center}
\includegraphics[width=0.45\textwidth,angle=0,clip=true]{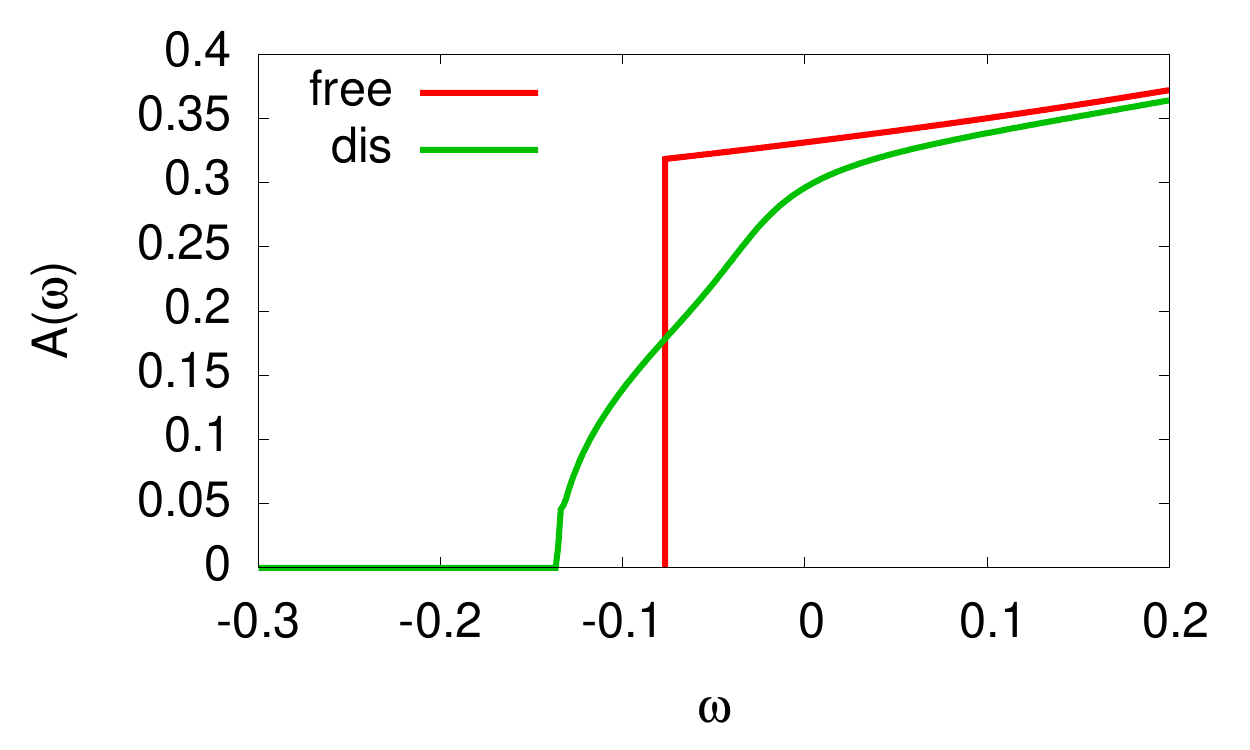}
\caption{DOS, $A(\omega)$, of the non-interacting system (free) shifted to match the 
filling of the $5\%$ doped system (dis). Unit of frequency is $D$; the zero of frequency 
is set to the chemical potential.}
\label{fig:DOSfree-dis}
\end{center}
\end{figure}

Here we present results obtained using basically two kinds of disorder. We first
consider a dichotomic disorder ($P_i$ distribution) in which a percentage $x=5\%$ of sites have a
lower energy $E_b=-0.5$ (in unit of the half bandwidth) than all the other
sites. This kind of disorder mimics the introduction of impurities associated
with doping. To this aim we fix the filling factor to the same value $x$. We
also consider gaussian uncorrelated disorder ($P_g$ distribution) which can mimic
a strong structural disorder, as usually happens in thin films. 
Even if $5\%$ of impurities seems to be a rather small quantity,
it can affect severely the lower part of the energy spectrum as can be seen in Fig. \ref{fig:DOSfree-dis}.
Moreover this is precisely the energy range in which electron-phonon interaction is relevant
($\omega\simeq\omega_0$).

\begin{figure*}[t!]
\centering
\includegraphics[width=1.0\textwidth,angle=0,clip=true]{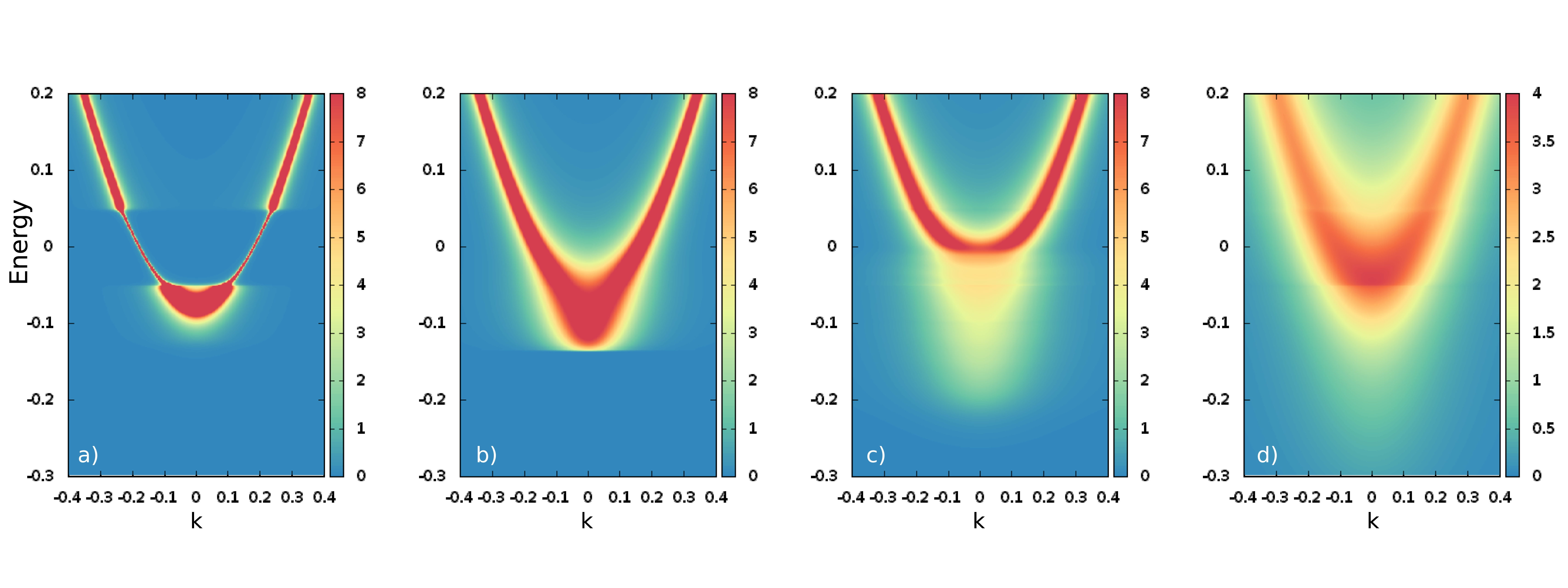}
\caption{The spectral function $A(k,\omega)$ for the LOC model.
a) Electron-phonon interaction only $\lambda=0.22$ 
b) Disorder only 
c) Electron-phonon interaction + disorder 
d) Electron-phonon interaction + gaussian disorder, the colourmap (range of z) has been expanded in this case to take into account
the lower value of the spectral function.}
\label{fig:spectra}
\end{figure*}

On top of this disordered system we consider a weak electron-phonon interaction
$\lambda=0.22$, which is the same in all the considered models. To
disentangle the separate action of electron-phonon and disorder we show the
spectral function in the case of LOC model in Fig. \ref{fig:spectra}. There the
spectral function is compared along a cut on $k_x$ axis around the $\Gamma$
point in the presence of electron-phonon interaction only (panel a)), in the
presence of impurities without electron-phonon interaction (panel b)) and under
the action of both electron-phonon and impurity-disorder in panel (c)). It's
immediately seen that the spectra in panel c) cannot be obtained by a simple
broadening of the spectra of panel a). A complete redistribution of the spectral
weight is obtained under the action of a quite low electron-phonon coupling in
presence of disorder. The growing of an impurity band appears to be evident at
the bottom of the coherent electronic band with a merging around the chemical
potential. On the other hand, the action of such a strong disorder does not
prevent the typical fingerprints of the electron-phonon interaction, as the kinks
at the phonon frequency (see Appendix \ref{AppendixD2Akw}). This result
highlights the fact that when disorder and electron-phonon coupling interact at
the same energy scales, as in the considered case, the action of disorder cannot
be taken into account as a simple broadening of the spectral features in absence
of disorder, since disorder and electron-phonon interaction work in a
cooperative way.

In panel d) we plot the spectra obtained using a gaussian disorder with
$\sigma^2=0.08$. We have chosen 
the variance of disorder requiring the same value of the Fermi $k_F$ as that given by the
$5$\% impurities. In this case an energy-dependent broadening can be seen in the
picture while the phonon signature, even weak, is still visible. Clearly the
interplay of impurities and distributed gaussian disorder with electron-phonon interaction
is very different.

\begin{figure}[htp]
\begin{center}
\includegraphics[width=0.45\textwidth,angle=0,clip=true]{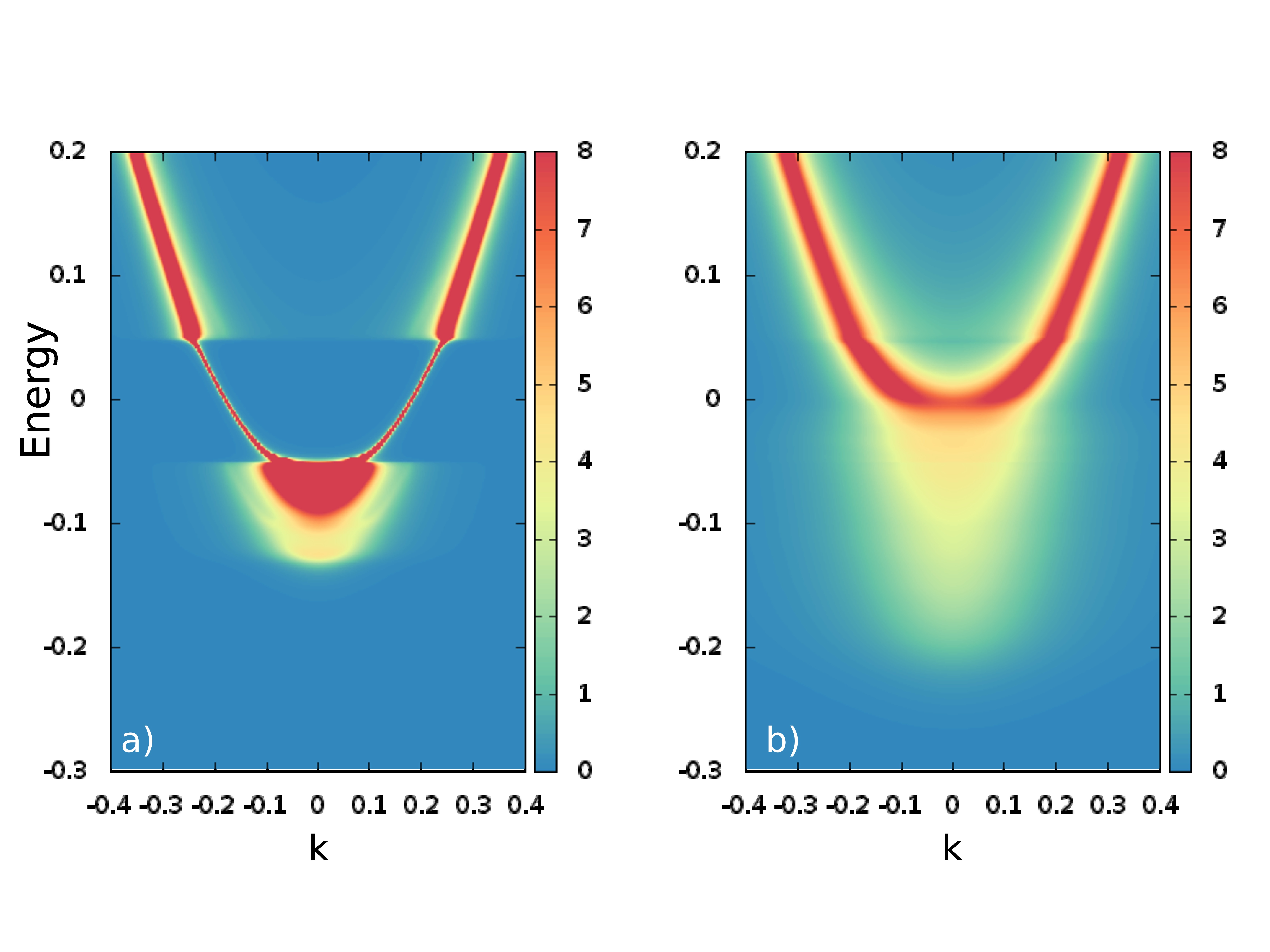}
\caption{The spectral function $A(k,\omega)$ for the NLOC model.
a) Electron-phonon interaction only $\lambda=0.22$ 
b) Electron-phonon interaction + disorder}
\label{fig:NLOCspectra}
\end{center}
\end{figure}

The scenario presented in Fig. \ref{fig:spectra} is rather general; indeed it
holds also in the case of highly non-local electron-phonon interaction. In Fig.
\ref{fig:NLOCspectra} we have considered an electron-phonon interaction of the
kind of Eq. (\ref{eq:Gvect}) with the screening $k$-vector $\kappa=10^{-3}$.
Comparing the spectra in absence of disorder (Fig. \ref{fig:spectra} a) and
\ref{fig:NLOCspectra} a)) we see that the enhanced forward scattering present in the
NLOC model broadens the low-energy features around the $\Gamma$ point. However
in the presence of impurities (Fig. \ref{fig:spectra} c) and
\ref{fig:NLOCspectra} b)) the spectra look much more similar even if phonon
signatures are more marked in the NLOC model. This is consistent with the
relevance of such a strong disorder at the highest binding energies. Increasing
the screening, the range of electron-phonon interaction reduces, and the
qualitative scenario becomes increasingly similar to that of LOC model. With the
chosen values of parameters at $\kappa=10^{-2}$ the spectra are almost
indistinguishable from those of Fig. \ref{fig:spectra}.

\begin{figure}[t]
\begin{center}
\includegraphics[width=0.45\textwidth,angle=0,clip=true]{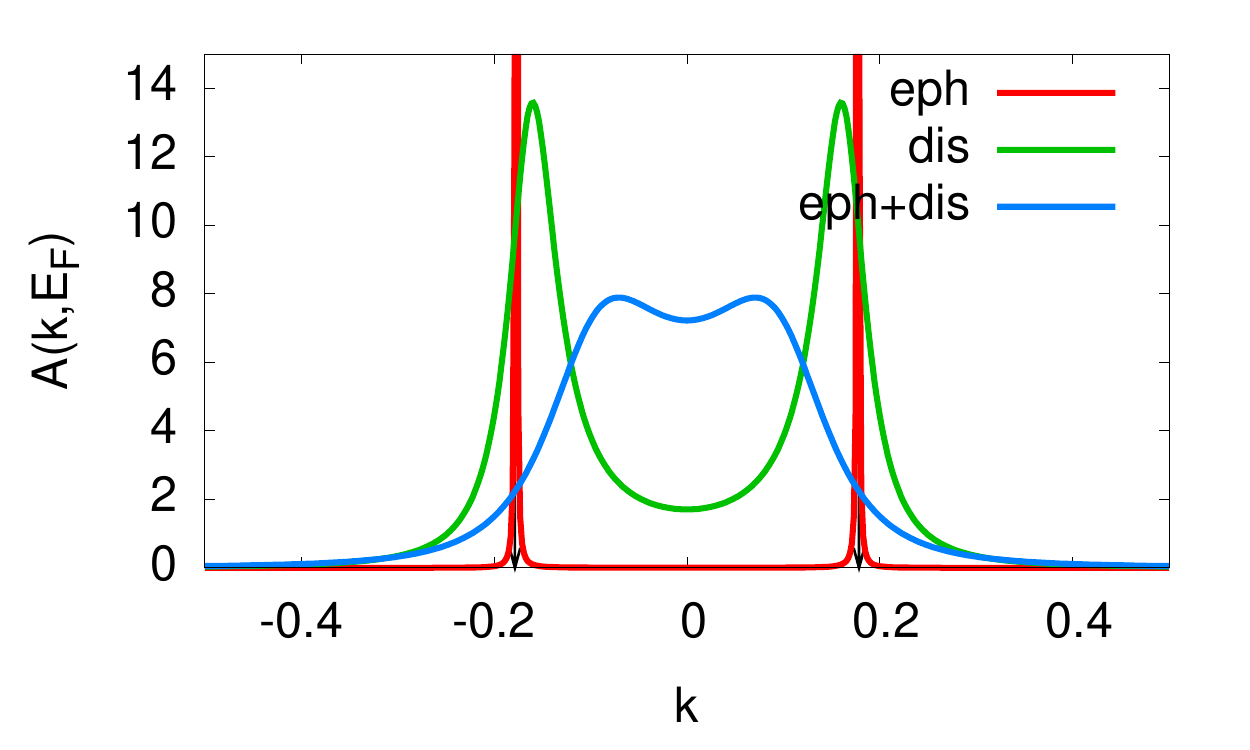}
\includegraphics[width=0.45\textwidth,angle=0,clip=true]{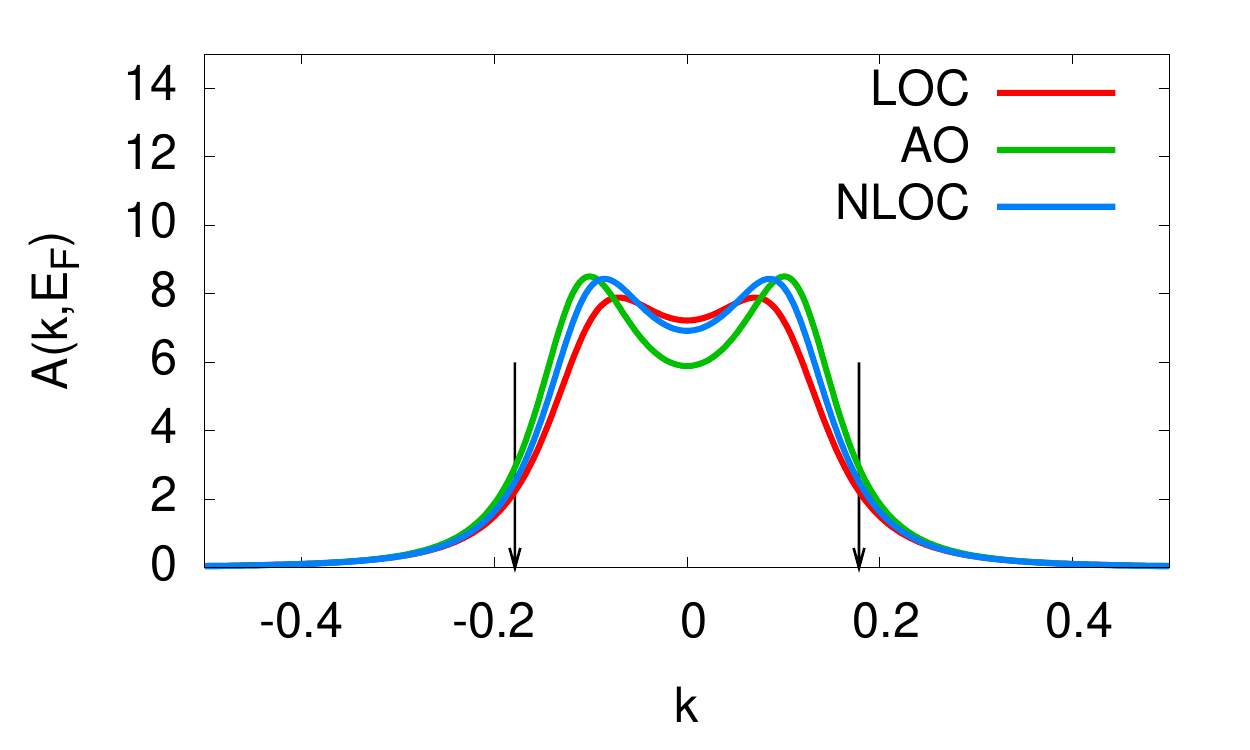}
\caption{Upper Panel: an MDC scan at Fermi energy in the LOC model. (eph) stands
for the non-disordered system under the action of electron-phonon interaction
only. (dis) is the purely disordered system without electron-phonon interaction.
(eph+dis) is the system under the action of both electron-phonon and disorder.
Lower Panel: an MDC scan at Fermi energy in the LOC compared with NLOC and AO model for
the same value of electron-phonon coupling $\lambda=0.22$ and the same disorder
variables $x=0.05,E_b=-0.5$. Vertical arrows mark the Luttinger's theorem value for $k_F$.}
\label{fig:Luttinger_x=0.05}
\end{center}
\end{figure}

A quantitative measure of the interplay between electron-phonon and disorder
effects can be probed by measuring the deviation of the Fermi wave-vector
($k_F$) from that predicted by Luttinger's theorem \cite{Luttinger-note} at a
given electron density. In Fig. \ref{fig:Luttinger_x=0.05} (upper panel) the
momentum distribution curve (MDC) is obtained from the spectral function. The
Luttinger's prediction for $k_F$ coincides with the position of the peaks in the
presence of electron-phonon interaction only. Indeed in this case the damping at
the Fermi energy is zero and the Fermi surface area is conserved; thus the sole
presence of electron-phonon interaction does not lead to a Fermi vector
reduction. Disorder alone, even strong as in our case, contributes to a
decreasing of $k_F$ only by $10\%$, while the additional presence of a
relatively weak electron-phonon interaction dramatically reduces $k_F$ by
$60\%$. If one takes the Luttinger's theorem \cite{Luttinger-note} for granted
in this conditions, the obtained electron density is far from the nominal one
given by the impurities' concentration. These evidences should be carefully
taken into account for the interpretation of experimental ARPES spectra, being
the fingerprint of a strong interplay between disorder and electron-phonon
interaction \cite{KyleSr2TiO4}. In the lower panel of Fig. \ref{fig:Luttinger_x=0.05} is shown a
comparison of the MDC curves for the LOC, NLOC and AO models. We see that the
reduction of $k_F$ is less effective in NLOC and AO models compared to LOC one.
We will discuss the reason for this behaviour below.

\begin{figure}[t]
\begin{center}
\includegraphics[width=0.5\textwidth,angle=0,clip=true]{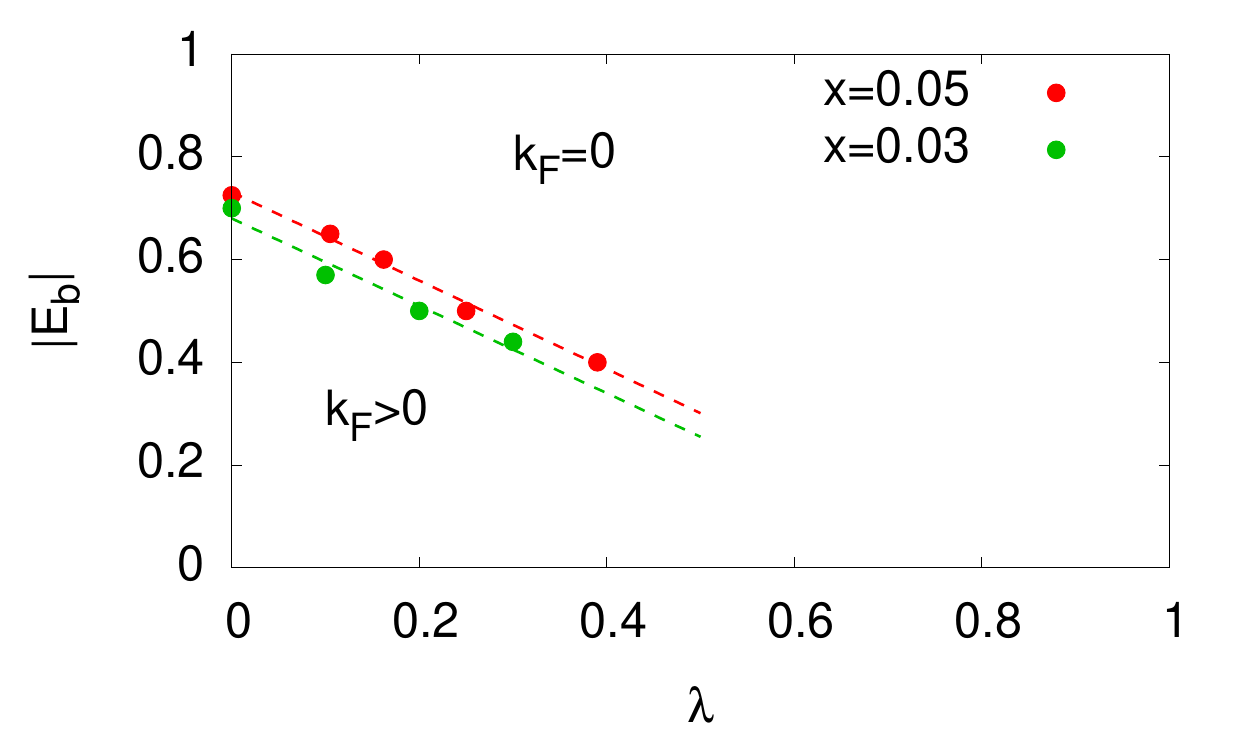}
\caption{The phase diagram of the LOC model at zero temperature for $x=0.03$ ans $x=0.05$.
Points are obtained at values of parameters such that $k_F=0$. 
Dashed lines are linear fits of the data. At a given value of $\lambda$ the increase of impurity 
concentration stabilizes the conductive phase.} 
\label{fig:Mott_phasediagram}
\end{center}
\end{figure}

The cooperative action of electron-phonon and disorder interactions is
particularly evident in the disorder-induced metal-insulator transition that occurs
as a function of the electron-phonon coupling $\lambda$. In this work, the disorder-induced 
MIT is defined looking at the vanishing of the Fermi vector $k_F$.
A vanishing  $k_F$ is a precursor of a 
vanishing density of states at the Fermi level, which in turn leads to an insulating state.
It is well known that, in a
disordered system, increasing the binding energy of the impurities will produce
a metal-insulator transition in which an impurity band detaches from the conduction band
\cite{Mott-note}.
Here we achieve the same phenomenon using the synergistic action of
electron-phonon interaction as it is shown in Fig. \ref{fig:Mott_phasediagram}
for two different impurity concentrations.

\begin{figure}[t]
\begin{center}
\includegraphics[width=0.45\textwidth,angle=0,clip=true]{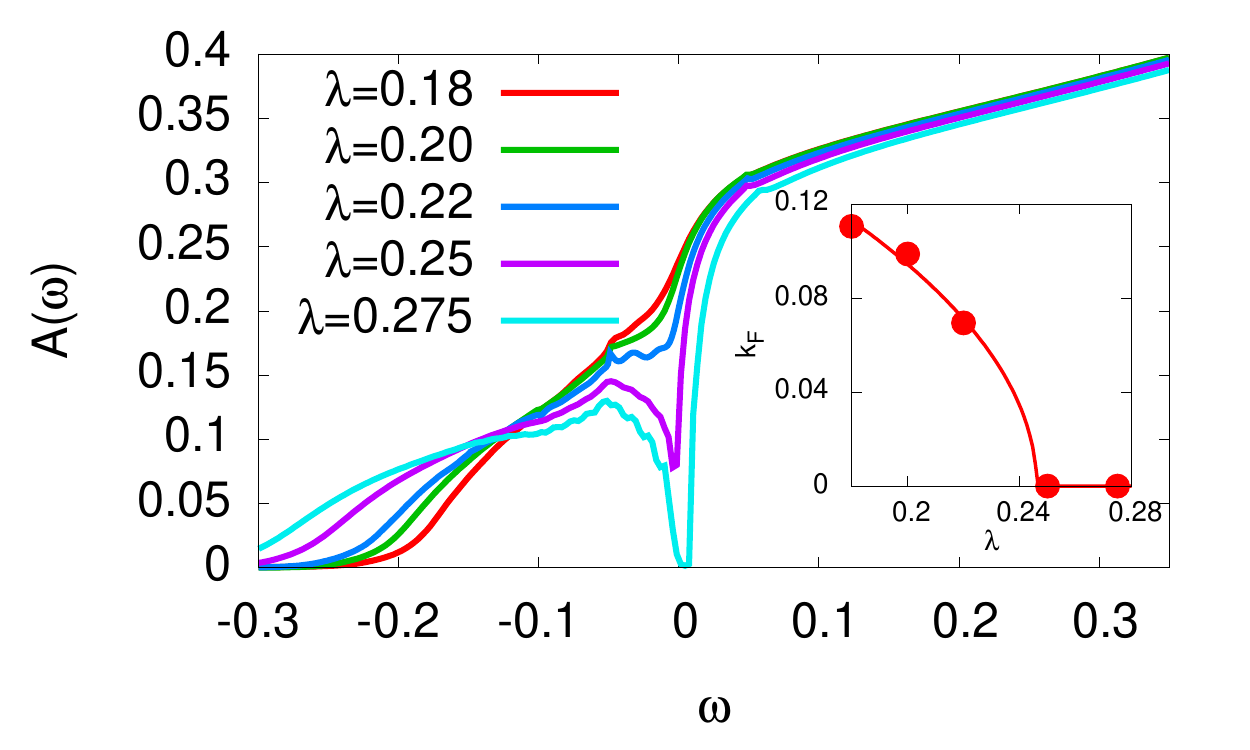}
\includegraphics[width=0.45\textwidth,angle=0,clip=true]{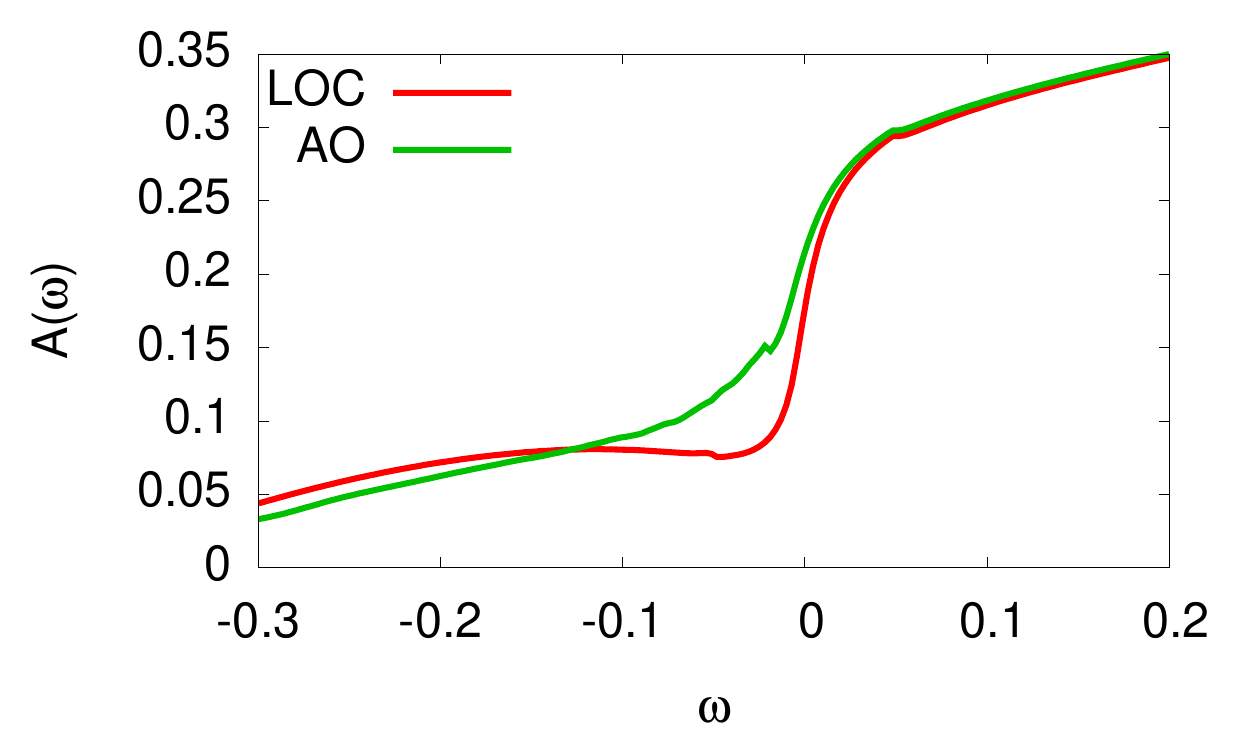}
\caption{Upper panel: the interacting density of the states for $x=0.05$ and
disorder level $E_b=-0.5$ as a function of electron-phonon coupling $\lambda$. In the inset is 
shown the value of $k_F$ as a function of $\lambda$.  
Lower panel: The DOS of the LOC and AO models at $\lambda=0.3$, here a gaussian
disorder of std. deviation $\sigma=0.05$ has been added to the dichotomic
disorder.} 
\label{fig:Mott_x=0.05}
\end{center}
\end{figure}

For a given value of $E_b=-0.5$  we report the DOS which clearly opens a gap at
$\lambda=0.275$ in Fig. \ref{fig:Mott_x=0.05} (upper panel). 
The vanishing of the Fermi surface occurs at a lower value of $\lambda$ as it is shown in the inset of the same figure.
The synergistic
work of electron-phonon interaction originates from the action of the Hartree
term Eq. (\ref{eq:Hartree}) which provides an electron-phonon induced increasing
of the binding energy which is proportional to the carrier density at a given
site. This is correlated with the presence of the impurity since the density
will be higher just at the impurity sites (see Appendix \ref{AppendixMott}). 
When electron-phonon-interaction is
non-local this effect is less marked as can be seen in Fig.
\ref{fig:Mott_x=0.05} (lower panel). For instance in the AO model, as the Hartree energy Eq.
(\ref{eq:NCABMMatsu}) does depend on the density on nearest neighbor planes
along the chain, the interplay between electron-phonon interaction and disorder
is less effective, as seen also in the smaller reduction of the Fermi surface
with respect to the LOC model (see Fig. \ref{fig:Luttinger_x=0.05} lower
panel).

\begin{figure}[h]
\begin{center}
\includegraphics[width=0.45\textwidth,angle=0,clip=true]{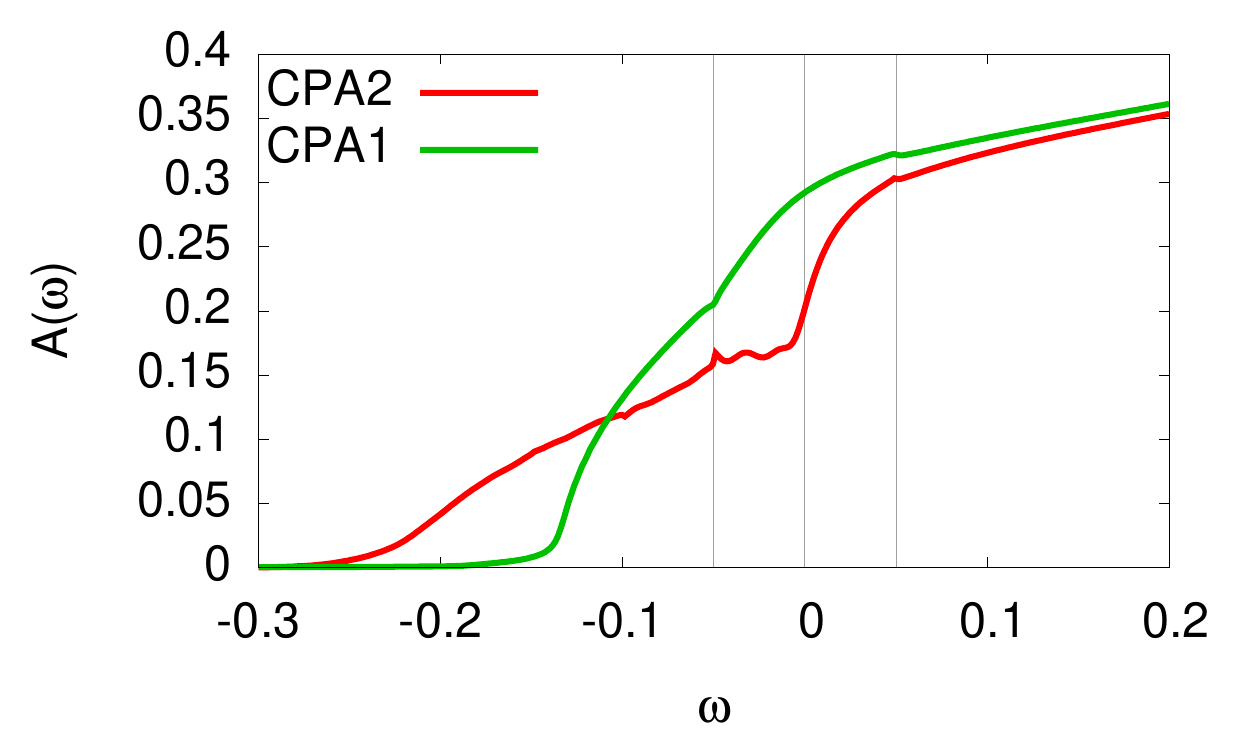}
\includegraphics[width=0.45\textwidth,angle=0,clip=true]{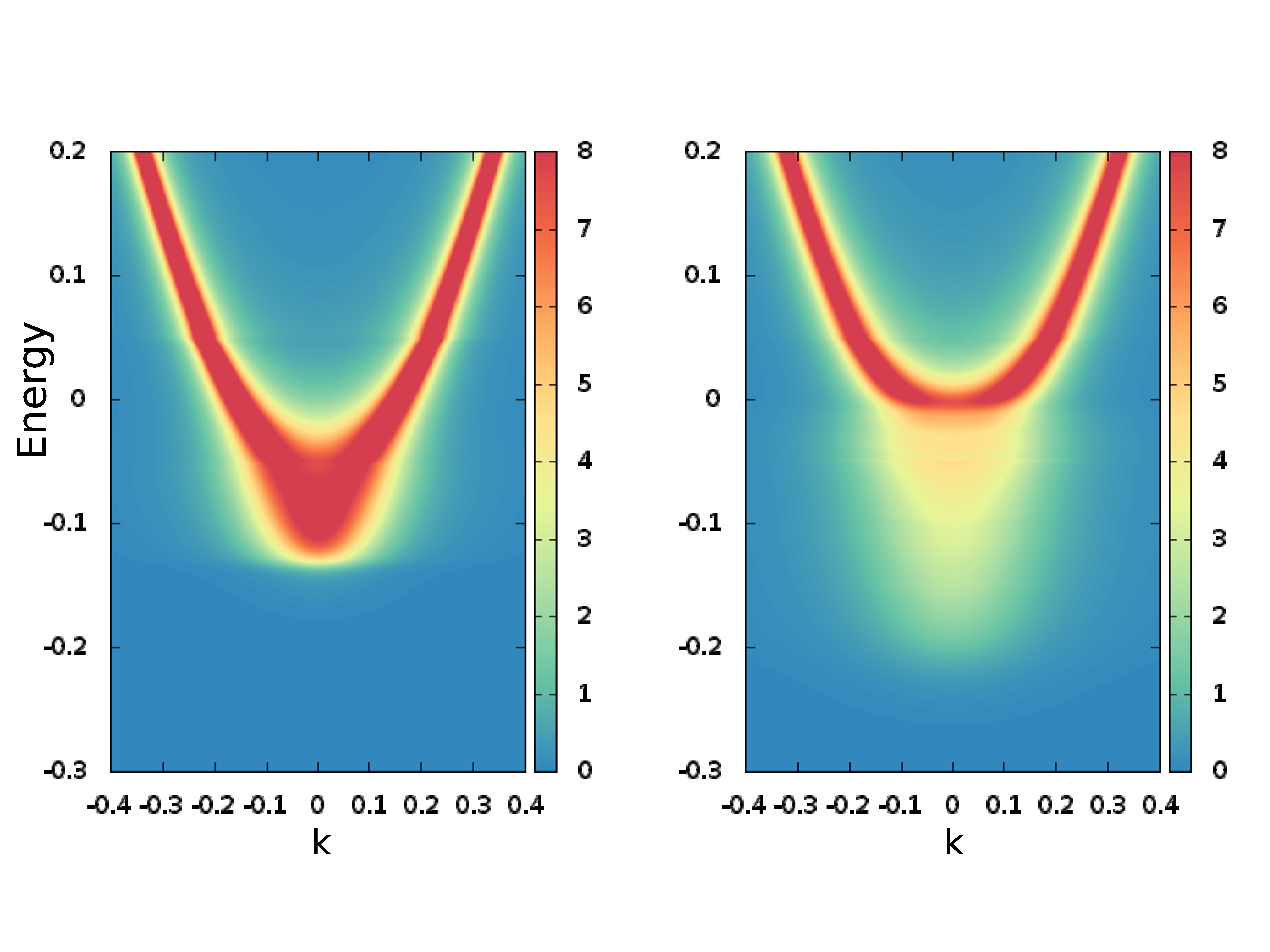}
\caption{Upper panel: DOS of LOC model within CPA1 and CPA2 approximations,vertical dotted lines marks the Fermi energy ($\omega=0$)
and the two phonon resonances at $\pm \omega_0$.
Lower panel: Comparison of CPA1 (left) and CPA2 (right) spectra.}
\label{fig:CPA2vsCPA1DOS}
\end{center}
\end{figure}

Moreover, a further insight into the interplay between electron-phonon and
disorder interaction can be obtained by the comparison of our results within the
two CPA schemes (see Section \ref{Methods}). The DOSs and the spectra obtained
by CPA1 and CPA2 approximations are compared in Fig. \ref{fig:CPA2vsCPA1DOS}
(upper and lower panels respectively). We see how the interplay between e-ph
interaction and disorder affect the DOS below the Fermi energy, just in the
energy region in which both disorder and e-ph are present. Noticeably phonon
signatures appear much more evident in the CPA2 scheme, and a large spectral
weight redistribution occurs at higher binding energies. Moreover we see that
within CPA1 scheme the effect of disorder is largely dominant, as can be seen by
comparing the spectrum of Fig. \ref{fig:CPA2vsCPA1DOS} (lower left panel) and
that obtained in the presence of pure disorder (see Fig. \ref{fig:spectra} c)).
Since in CPA1 we average the electron-phonon self-energy over the disorder
variable we can ascribe the large discrepancies between the spectra in Fig.
\ref{fig:CPA2vsCPA1DOS} to the correlation between electron-phonon and disorder
effects in the self-energy. This issue can be analyzed from the point of view of
perturbative expansions. The resummation in CPA2 scheme of diagrams of type a)
in Fig. \ref{fig:CPA2vsCPA1diagrams} which take into account the correlation at
the Hartree level between electron-phonon and local disorder, leads to an
enhancement of the electron-phonon interaction effects on the energy scale of
the emerging impurity band (around $\simeq E_b$ from the Fermi level). In
contrast to CPA1 the CPA2 Hartree term is correlated to the presence of the
impurity leading to the $\lambda$ dependence of the disorder-induced metal-insulator transition (see
discussion above and upper panel of Fig. \ref{fig:Mott_x=0.05}). For this reason,
as shown in Fig. \ref{fig:CPA2vsCPA1DOS}, the impurity band within CPA2 seems to
be more marked than that in CPA1. However another aspect is clear from the
comparison in Fig. \ref{fig:CPA2vsCPA1DOS}: the CPA2 impurity band is also much
wider that that obtained within CPA1, and, despite the strong disorder, prominent
phonon signatures are still evident on the impurity band. This should be
ascribed to the correlation between disorder and electron-phonon self-energy at
the Fock level diagrams of type b) in Fig. \ref{fig:CPA2vsCPA1diagrams}.
In previous work an interplay between electron-phonon interaction and disorder has been found
within self-consistent Born approximation
\cite{BftPRB2003,KnigavkoCarbottePRB2006,DoganMarsiglioJSupMagn2007} in which,
despite the self-energy separates into  electron-phonon and disorder parts, 
non additivity in electron scattering time is found due to the
self-consistency condition. We
remark here the difference in our strong-disorder approach in which the
self-energy appearing in Eq. (\ref{eq:Gloc}) is no longer separable into two
contributions. We thus have analyzed the strong fluctuations of the self-energy due to
disorder rather than its separability into electron-phonon and disorder part.

%

\section{Conclusions}
\label{Conclusions}
In conclusion, in this work we have investigated the role of the electron-phonon
interaction in disordered systems, and their strong interplay when the energy
scales in which they act are comparable. It is well known that trapping
impurities provide the necessary energy for the polaronic transition stabilizing
the polaronic state at weaker electron-phonon coupling
\cite{Alexandrov,Cataudella,Mishchenko,Berciu}. Here we have discussed this
interplay at finite electron density and weak electron-phonon coupling, thus
relying our study on the PPNCA to deal with weak electron-phonon interaction. We have developed a
theoretical method to combine the PPNCA with the 
CPA to study strongly disordered systems, and we have extended such theory to
the Apical Oxygens model \cite{SawatzkyBM} and to a non-local electron-phonon
interaction characteristic of couplings with crystal's polar modes. We mainly
focused our attention on low dimensional systems such as quasi twodimensional or
layered ones, since in these cases the effect of disorder can in principle be
larger with respect to purely 3D systems.  On the other hand, we concentrated on low
doped systems in which the impurity band can be very close, and hybridizes, with
the bottom of the electronic one. This peculiar, but quite common experimental
and theoretical evidence \cite{OxygenVacancesSrTiO3, MeevasanaSrTiO3eph,
NbDopedSrTiO3, KyleSr2TiO4, Si-doped_Ga, OkabayashiARPESMnDopedSemicondPRB2001,
JarrelImpurityBandsMnGaAsPRB2006} allowed us to study when disorder and
electron-phonon interaction act in a cooperative way, and the action of disorder
cannot be included in a perturbative way as a source of weak broadening of the
spectral features. On the contrary, impurity-type disorder strongly affects the
electronic structure giving rise to a significant spectral weight
redistribution. This could lead to a dramatic Fermi surface reduction even at
moderate electron-phonon couplings, which in turns can be detected as a
Luttinger's theorem violation\cite{KyleSr2TiO4} and eventually an electron-phonon driven 
metal-insulator transition as the Fermi surface vanishes. From a quantitative point of
view, the strongest interplay between electron-phonon and local disorder is
found for the local electron-phonon interaction (LOC model). Non-local couplings
studied in this work (AO, NLOC) both display a less  effective interplay with
disorder as a consequence of the interactions' non-locality.

CPA approximation used to approach the strong disorder regime is a reasonable
approximation for the DOS or the average spectral function in three dimensions \cite{DobroEPL2003}. 
In our 2 (LOC,NLOC) or 2+1 (AO) dimensional systems there are however some 
deviations which can be treated within 
a non-local DCA framework \cite{JarrelKrishnamurthyPRB2001}. Generally speaking CPA 
overestimates the disorder induced gap. As a consequence in our treatment of the disorder-induced MIT
we expect that the disorder needed to reach the MIT would be slightly lower going beyond CPA.
Localization effects, absent in CPA approach, which are however beyond the present work, can be relevant for transport
properties in low dimensional systems. Their effects can be probed at the local
level by anomalous fluctuations of the local DOSs which can be relevant to
tunneling experiments. Instead of the averaged DOS taken into
account in this work one can consider the typical DOS obtained as geometric
averages of local DOSs \cite{DobroEPL2003}. As far as local quantities are
concerned for the LOC model one can generalize our self-consistency equations to
the case of typical quantities following the lines of refs. \cite{DobroEPL2003,FehskeAndHol}. 

PPNCA approximation for electron-phonon interaction used in our work cannot be
used to attach the polaronic regime which can be interesting to study, since
the recently found polaronic resonances in single layer high-$T_c$
superconducting $FeSe$ \cite{ZXShen}. Also from a theoretical point of view, the
interplay between disorder and polaronic electron-phonon interaction could be
much different from that proposed in the present paper \cite{SawatzkyBM}. To
this aim a beyond-NCA approach such as DMFT should be useful
also to include electronic correlations. A cluster-DMFT approach could also be useful to include
spatial correlations which we neglect in our local approach in the LOC model case,
overcoming in this way the well known problems of single site DMFT in dealing with systems at low
dimensionality \cite{clusterDMFT}.

The authors kindly thank Prof. K. Shen and Dr. Y. Nie for very helpful
discussions and hints on the work. Helpful discussions with Dr. E. Cappelluti,
Dr. P. Barone, Prof. G. Sangiovanni and Prof. R. Valent\'{i} are also acknowledged. The computational
support from the CINECA supercomputing center under the grant "ConvR\_aq\_caspF"
is gratefully acknowledged.


\begin{thebibliography}{99}

\expandafter\ifx\csname natexlab\endcsname\relax\def\natexlab#1{#1}\fi
\expandafter\ifx\csname bibnamefont\endcsname\relax
  \def\bibnamefont#1{#1}\fi
\expandafter\ifx\csname bibfnamefont\endcsname\relax
  \def\bibfnamefont#1{#1}\fi
\expandafter\ifx\csname citenamefont\endcsname\relax
  \def\citenamefont#1{#1}\fi
\expandafter\ifx\csname url\endcsname\relax
  \def\url#1{\texttt{#1}}\fi
\expandafter\ifx\csname urlprefix\endcsname\relax\def\urlprefix{URL }\fi
\providecommand{\bibinfo}[2]{#2}
\providecommand{\eprint}[2][]{\url{#2}}


\bibitem{Damascelli} A. Damascelli, Z, Hussain and Z. X. Shen, Rev. Mod. Phys. {\bf 75}, 473 (2003)

\bibitem{Nagaosa} T. Cuk, D. H. Lu, X. J. Zhou, Z.-X. Shen, T. P. Devereaux, and N. Nagaosa,  phys. stat. sol. (b) {\bf 242}, 11 (2005)

\bibitem{Dagotto} E. Dagotto, T. Hotta and A. Moreo, Physics Report {\bf 334}, 1 (2001)

\bibitem{Tokura} Y. Tokura, Physics Today, July 2003, pp 50-55.

\bibitem{Hasan} M. Z. Hasan and C. L. Kane, Rev. Mod. Phys. {\bf 82}, 3045 (2010)

\bibitem{CastroNeto} A. H. Castro Neto, F. Guinea, N. M. R. Peres, K. S. Novoselov and A. K. Geim, Rev. Mod. Phys. {\bf 81}, 109 (2009)

\bibitem{OxygenVacancesSrTiO3} Young Jun Chang, Aaron Bostwick, Yong Su Kim,
Karsten Horn, and Eli Rotenberg Phys. Rev. B {\bf 81}, 235109(2010)

\bibitem{MeevasanaSrTiO3eph} W Meevasana, X J Zhou, B Moritz, C-C Chen, R H He, S-I Fujimori, D H Lu, S-K Mo, R G Moore, F Baumberger, T P Devereaux, D van der Marel, N Nagaosa, J Zaanen and Z-X Shen, New J. of Phys. {\bf 12}, 023004 (2010) 

\bibitem{NbDopedSrTiO3} M. Takizawa, K. Maekawa, H. Wadati, T. Yoshida, and A. Fujimori
H. Kumigashira and M. Oshima, Phys. Rev. B {\bf 79}, 113103 (2009)

\bibitem{ZXShen} J. J. Lee, F. T. Schmitt, R. G. Moore, S. Johnston, Y.-T. Cui,
W. Li, M. Yi, Z. K. Liu, M. Hashimoto, Y. Zhang, D. H. Lu, T. P. Devereaux, D.
-H. Lee, Z.-X. Shen arXiv:1312.2633

\bibitem{Anatase} S. Moser, L. Moreschini, J. Ja\'cimovi\'c, O. S. Bari\v{s}i\'c, H. Berger, A. Magrez, Y. J. Chang, K. S. Kim, A. Bostwick, E. Rotenberg, L. Forr\'o, and M. Grioni
Phys. Rev. Lett. {\bf 110}, 196403 (2013)

\bibitem{BaKBiO3} R. Nourafkan, F. Marsiglio and G. Kotliar, Phys. Rev. Lett. {\bf 109}, 017001 (2012)  

\bibitem{eph-TI} T. Kondo, Y. Nakashima, Y. Ota, Y. Ishida, W. Malaeb, K. Okazaki, S. Shin,  M. Kriener, S. Sasaki, K. Segawa and Y. Ando, Phys. Rev. Lett. {\bf 110}, 217601 (2013)

\bibitem{KyleSr2TiO4}  Y.F. Nie, D. Di Sante, S. Chatterjee, P.D.C. King, M. Uchida, S. Ciuchi, D.G. Schlom,  and K.M. Shen in preparation

\bibitem{Si-doped_Ga} P. Richard, T. Sato, S. Souma, K. Nakayama, H. W. Liu, K. Iwaya, T. Hitosugi, H. Aida, H. Ding, and T. Takahashi, App. Phys. Lett. {\bf 101}, 232105 (2012)

\bibitem{OkabayashiARPESMnDopedSemicondPRB2001} J. Okabayashi, A. Kimura, O.
Rader, T. Mizokawa, A. Fujimori, T. Hayashi, and M. Tanaka, Phys. Rev.  B, {\bf
64}, 125304 (2001)
 
\bibitem{JarrelImpurityBandsMnGaAsPRB2006} M. A. Majidi, J. Moreno, M. Jarrell,
R. S. Fishman, and K. Aryanpour, Phys. Rev.  B, {\bf 74}, 115205 (2006) 

\bibitem{Claessen} G. Berner, M. Sing, H. Fujiwara, A. Yasui, Y. Saitoh, A. Yamasaki, Y. Nishitani, A. Sekiyama, N. Pavlenko, T. Kopp, C. Richter, J. Mannhart, S. Suga, and R. Claessen, Phys. Rev. Lett {\bf 110}, 247601 (2013)

\bibitem{Santander_Syro} A. F. Santander-Syro, O. Copie, T. Kondo, F. Fortuna, S. Pailh\'{e}s, R. Weht, X. G. Qiu, F. Bertran, A. Nicolaou, A. Taleb-Ibrahimi, P. Le F\'{e}vre, G. Herranz, M. Bibes, N. Reyren, Y. Apertet, P. Lecoeur, A. Barth\'{e}l\'{e}my and M. J. Rozenberg, Nature {\bf 469}, 189 (2011)

\bibitem{Valenti} J. Shen, H. Lee, R. Valent\'{i} and H. O. Jeschke, Phys. Rev. B {\bf 86}, 195119 (2012)

\bibitem{Reyren} N. Reyren, S. Thiel, A. D. Caviglia, L. Fitting Kourkoutis, G. Hammerl, C. Richter, C. W. Schneider, T. Kopp, A.-S. R\"{u}etschi, D. Jaccard, M. Gabay, D. A. Muller, J.-M. Triscone, and J. Mannhart, Science {\bf 317}, 1196 (2007)

\bibitem{Das_Sil} A. N. Das and S. Sil, Physics Letters A {\bf 348}, 266 (2006) 

\bibitem{Alexandrov} J. P. Hague, P. E. Kornilovitch, and A. S. Alexandrov, Phys. Rev. B  {\bf 78}, 092302 (2008)

\bibitem{Mishchenko} M. Berciu, A. S. Mishchenko and N. Nagaosa, Eur. Phys. Lett. {\bf 89}, 37007 (2010) 

\bibitem{Cataudella}  C. A. Perroni and V. Cataudella, Phys. Rev. B  {\bf 85}, 155205 (2012)

\bibitem{Berciu} Hadi Ebrahimnejad and Mona Berciu, Phys. Rev. B  {\bf 85}, 165117 (2012)

\bibitem{BftPRB2003} E. Cappelluti and L. Pietronero, Phys. Rev. B {\bf 68}, 224511 (2003)

\bibitem{KnigavkoCarbottePRB2006} A. Knigavko, J. P. Carbotte, Phys. Rev. B {\bf 73}, 125114 (2006)

\bibitem{DoganMarsiglioJSupMagn2007} F. Dogan and F. Marsiglio, J. Sup. Nov. Magn. {\bf 20}, 225 (2007)

\bibitem{ChenPRB1972} A. Chen, G. Weisz, A. Scher, Phys. Rev. B {\bf 5}, 2897 (1972)

\bibitem{GirvinPRB1980} S.M. Girvin, M. Jonson, Phys. Rev. B {\bf 22}, (1980)

\bibitem{LetfulovFreericksPRB2002} B. M. Letfulov, J. K. Freericks, Phys. Rev. B {\bf 66}, 033102 (2002)

\bibitem{FreericksRMP2003} J. K. Freericks and V. Zlati\'c Rev. Mod. Phys. {\bf 75}, 1333 (2003)

\bibitem{note-disorder} It is worth to note that even a modest amount of impurities can introduce
strong disorder effects at the energy scale of the order of the impurities' binding energy.

\bibitem{DasSarma} S. Das Sarma and B. A. Mason, Annals of Physics {\bf 163}, 78 (1985)

\bibitem{BMcuprates} O. Gunnarsson and O. R\"{o}sch, J. Phys.: Condens. Matter {\bf 20}, 043201 (2008)

\bibitem{SawatzkyBM} Bayo Lau, Mona Berciu, and George A. Sawatzky Phys. Rev. B {\bf 76}, 174305 (2007)

\bibitem{Siggia} L. Schwartz and E. Siggia, Phys. Rev. B {\bf 5}, 383 (1972)

\bibitem{Vollhardt} D. Vollhardt, in Correlated Electron Systems, edited by V. J.
Emery (World Scientific, Singapore, 1992).

\bibitem{DMFT} A. Georges, G. Kotliar, W. Krauth, and M.J. Rozenberg, Rev. Mod. Phys. {\bf 68}, 13 (1996)

\bibitem{Bronold} F. X. Bronold, A. Saxena and A. R. Bishop,  Phys. Rev. B {\bf 63}, 235109 (2001)

\bibitem{note-NCA} To avoid confusion with the acronym NCA used in the theory of strongly correlated electrons system,
we name our approximation PPNCA to indicate a theory which includes all non-crossing diagrams with respect to phonon
propagator. Notice that our theory includes diagrams in which disorder insertions cross with phonon propagators
(see Fig. \ref{fig:CPA2vsCPA1diagrams})

\bibitem{virtual} D. Stroud and H. Ehrenreich, Phys. Rev. B {\bf 2}, 3197 (1970)

\bibitem{Luttinger-note} In two dimensions, for a circular Fermi surface, the expression for $k_F$ is
 $k_F=\sqrt{2 x/\pi}$ (in units of ($\pi/a$)).

\bibitem{Mott-note} With our choose of DOS at $x=0.05$ CPA gives $E_b=-0.725$ for the disorder-induced Mott transition.

\bibitem{DobroEPL2003} V. Dobrosavljevi\'c, A. A. Pastor and B. K. Nikoli\'c, Europhys. Lett. {\bf 62}, 76 (2003)

\bibitem{JarrelKrishnamurthyPRB2001} M. Jarrell and H. R. Krishnamurthy, Phys. Rev. B {\bf 63}, 125102 (2001)

\bibitem{FehskeAndHol} H. Fehske, F.-X. Bronold, and A. Alvermann, Proc.
Int. School of Physics ``Enrico Fermi'', Course CLXI, Polarons in Bulk Materials
and Systems with Reduced Dimensionality, Eds. G. Iadonisi, J. Ranninger, G. de
Filippis, IOS Press, Amsterdam, Oxford, Tokio, Washington DC, 313 (2006);
F.-X. Bronold,  A. Alvermann and H. Fehske,  Phil. Mag. {\bf 84}, 673 (2004)

\bibitem{clusterDMFT} T. Maier, M. Jarrell, T. Pruschke and M. Hettler, Rev. Mod. Phys. {\bf 77}, 1027 (2005)

\end{thebibliography}

\appendix
\section{Second derivative of the spectral function}
\label{AppendixD2Akw}

\begin{figure}[h]
\begin{center}
\includegraphics[width=0.45\textwidth,angle=0,clip=true]{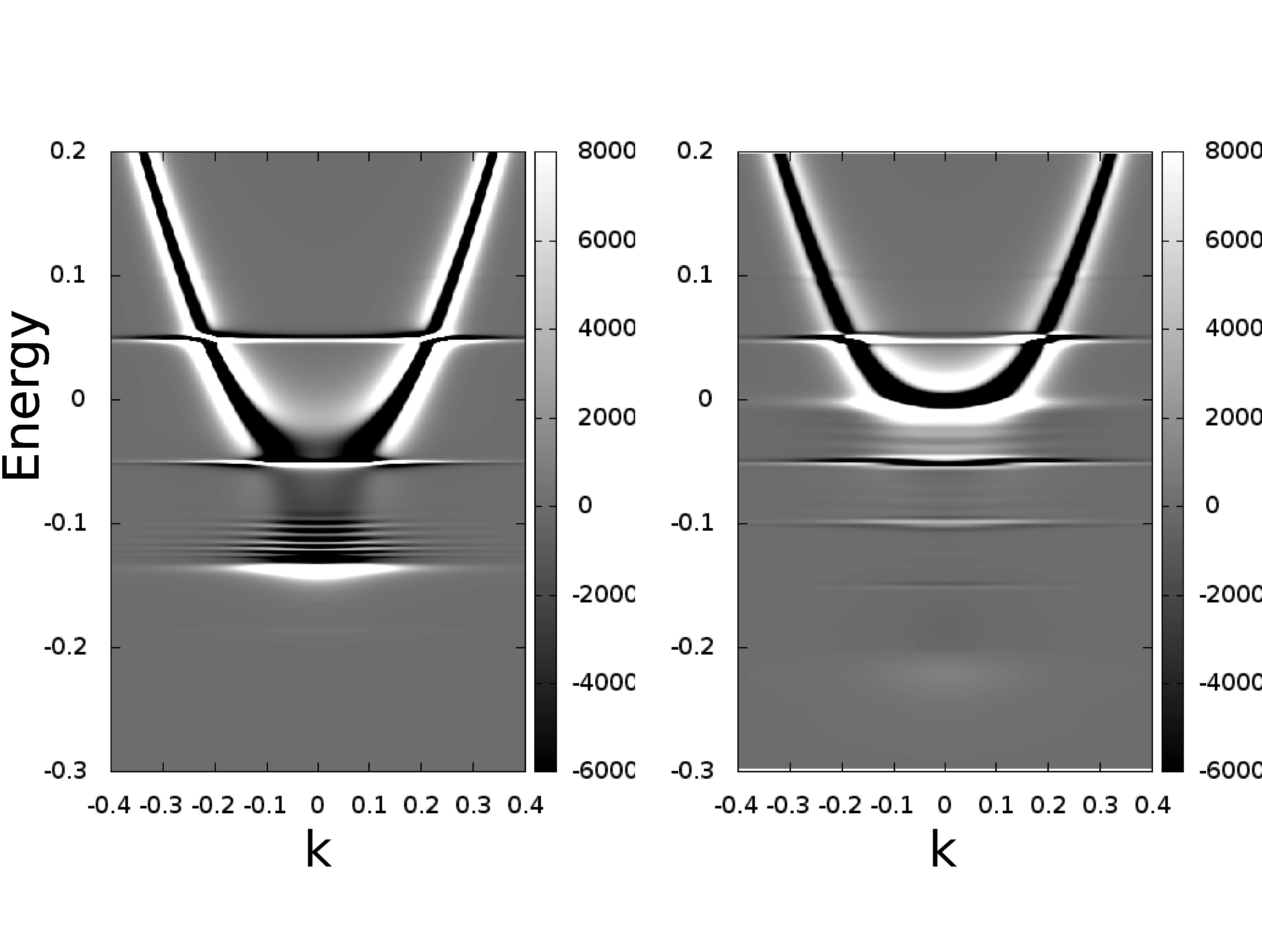}
\caption{Comparison of second derivative of the spectral function within CPA1 (left) and CPA2 (right) spectra.}
\label{fig:CPA2vsCPA1DOSD2Akw}
\end{center}
\end{figure}

A common used technique to highlight subtle spectral features is to take the second derivative of the spectral function
$\frac{\partial^2}{\partial \omega^2}A({\bf k},\omega)$. In Fig. \ref{fig:CPA2vsCPA1DOSD2Akw} we plot this function
using CPA1 and CPA2 iteration schemes. In both cases the phonon's signatures are evident but a little bit more within CPA2.
More importantly at higher binding energies, CPA1 spectra clearly shows disorder non-dispersed features while in CPA2 clear phonon's 
higher order resonances are visible up to fourth order, even in the presence of such a strong disorder.

\section{Electron-phonon induced Mott transition}
\label{AppendixMott}
Let us consider the bimodal disorder case $P_i(\xi)=
x\delta(\xi-E_b)+(1-x)\delta(\xi)$ in the LOC model. 
Let us consider only the action of the Hartree term
in the self-energy Eq. (\ref{eq:Hartree}), so that the single-site Green function Eq. (\ref{eq:Gimp}) reads
\begin{eqnarray}
\label{eq:GimpApp}
\mathcal{G}(\omega)=\frac{x}{G_{0}^{-1}(\omega)-E_b+\lambda n_1} + \frac{1-x}{G_{0}^{-1}(\omega)+\lambda n_0},
\end{eqnarray}
where
\begin{eqnarray}
n_1= -\frac{1}{\beta}\sum_n \frac{x}{G_{0}^{-1}(\omega)-E_b+\lambda n_1}e^{i\omega_n 0^+}\label{eq:n0App}\\
n_0= -\frac{1}{\beta}\sum_n \frac{1-x}{G_{0}^{-1}(\omega)+\lambda n_0}e^{i\omega_n 0^+}\label{eq:n1App}\quad,
\end{eqnarray}
where $n_1$ is the electron density in the impurity site and $n_0$ is the
density everywhere else. In the atomic (zero hopping) limit we have
$n_1=1,n_0=0$ but due to the hibridization of the impurity sites $n_1<1$ and
$n_0>0$. From Eqs. (\ref{eq:GimpApp},\ref{eq:n0App},\ref{eq:n1App}) it is evident that as
far as the electron-phonon interaction is considered at the Hartree level
$E_b\rightarrow E_b-\lambda(n_1-n_0)$ and the disorder-induced metal-insulator transition occurs when
\begin{equation}
\label{eq:MottApp}
|E_b|=|E_{MIT}|-\lambda(n_1-n_0)
\end{equation}
with $|E_{MIT}|$ the binding energy at the impurity site needed to detach
the impurity band in absence of electron-phonon interaction. Eq.
(\ref{eq:MottApp}) explains the linear dependence found for small $\lambda$ for
the disorder-induced metal-insulator transition in Fig. \ref{fig:Mott_phasediagram}. 
It is worth to note that this effect is absent in 
CPA1 where the electron-phonon self-energy is mediated and as a consequence there is no 
electron-phonon contribution to the binding energy at the impurity site.

\end{document}